# Temperature-dependent refractive index of CaF$_2$ and Infrasil 301

Douglas B. Leviton[*], Bradley J. Frey, Timothy J. Madison

NASA Goddard Space Flight Center, Greenbelt, MD 20771

**ABSTRACT**

In order to enable high quality lens designs using calcium fluoride (CaF$_2$) and Heraeus Infrasil 301 (Infrasil) for cryogenic operating temperatures, we have measured the absolute refractive index of these two materials as a function of both wavelength and temperature using the Cryogenic, High-Accuracy Refraction Measuring System (CHARMS) at NASA's Goddard Space Flight Center. For CaF$_2$, we report absolute refractive index and thermo-optic coefficient (dn/dT) at temperatures ranging from 25 to 300 K at wavelengths from 0.4 to 5.6 μm, while for Infrasil, we cover temperatures ranging from 35 to 300 K and wavelengths from 0.4 to 3.6 μm. For CaF$_2$, we compare our index measurements to measurements of other investigators. For Infrasil, we compare our measurements to the material manufacturer's data at room temperature and to cryogenic measurements for fused silica from previous investigations including one of our own. Finally, we provide temperature-dependent Sellmeier coefficients based on our measured data to allow accurate interpolation of index to other wavelengths and temperatures.

## 1. INTRODUCTION

Over the past few years, using the Cryogenic High Accuracy Refraction Measuring System (CHARMS)[1-2] at GSFC, we have measured the absolute refractive indices of several visible and infrared optical materials with very high accuracy down to temperatures as low as 15 K using the method of minimum deviation refractometry.[3-7] The present study reports measurements of one sample each of single crystal calcium fluoride (CaF$_2$) and Infrasil 301, a formulation of synthetic fused silica manufactured by Hereaus Quarzglas optimized for transmission in the near infrared wavelength region. Measurements for these two relatively low index materials are considered uncertain at 1 or 2 parts in the fifth decimal place of index over the wavelength range from 0.4 to 5.6 microns CaF$_2$ from 35 to 300 K and from 0.4 to 3.6 microns for Infrasil from 35 to 300 K.

The prisms for this study were fabricated to fit in the custom, windowless, CHARMS sample chamber and to provide optimal deviation angles for measurements of highest precision and lowest uncertainty. The apex angle of the prism is designed so that beam deviation angle for the highest index in the material's transparent range will equal the largest accessible deviation angle of the refractometer, 60°. The nominal apex angles for the prisms measured in this study are: CaF$_2$ – 54.0°; Infrasil 301 – 58.0°. Measurements of Infrasil are compared to our previous cryogenic measurements of Corning 7980 synthetic fused silica, while measurements of CaF$_2$ are compared to copious room temperature and sparse cryogenic measurements in the literature. Interspecimen variability studies of CaF$_2$ in our measurement facility are planned.

## 2. TREATMENT OF MEASURED INDEX DATA

Detailed descriptions of our data acquisition and reduction processes are documented elsewhere[8], as are our calibration procedures[9]. In general we fit our raw measured data to a wavelength and temperature dependent Sellmeier model of the form:

$$n^2(\lambda, T) - 1 = \sum_{i=1}^{m} \frac{S_i(T) \cdot \lambda^2}{\lambda^2 - \lambda_i^2(T)}$$

---

[*] doug.leviton@nasa.gov, phone 1-301-286-3670, FAX 1-301-286-0204

where $S_i$ are the strengths of the resonance features in the material at wavelengths $\lambda_i$. When dealing with a wavelength interval between wavelengths of physical resonances in the material, the summation may be approximated by only a few terms, m – typically three[10]. In such an approximation, resonance strengths $S_i$ and wavelengths $\lambda_i$ no longer have direct physical significance but are rather parameters used to generate an adequately accurate fit to empirical data. If these parameters are assumed to be functions of T, one can generate a temperature-dependent Sellmeier model for $n(\lambda,T)$.

This modeling approach has historically worked well for a variety of materials despite rather scarce available index measurements – to cover a wide range of temperatures and wavelengths – upon which to base a model. One solution to the shortcoming of lack of measured index data has been to appeal to room temperature refractive index data at several wavelengths to anchor the model and then to extrapolate index values for other temperatures using accurate measurements of the thermo-optic coefficient dn/dT at those temperatures, which are much easier to make than accurate measurements of the index itself at exotic temperatures.[10] This is potentially quite risky depending on the sample material and required accuracy of refractive index knowledge, especially to temperatures well-separated from room temperature.

Meanwhile, with CHARMS, we measure index directly, densely sampling over a wide range of wavelengths and temperatures to produce a model with residuals on the order of the uncertainties in our raw index measurements. For our models, we have found that 4th order temperature dependences in all three terms in each of $S_i$ and $\lambda_i$ work adequately well, as also found previously in the literature.[10] The Sellmeier equation consequently becomes:

$$n^2(\lambda,T) - 1 = \sum_{i=1}^{3} \frac{S_i(T) \cdot \lambda^2}{\lambda^2 - \lambda_i^2(T)}$$

where,

$$S_i(T) = \sum_{j=0}^{4} S_{ij} \cdot T^j$$

$$\lambda_i(T) = \sum_{j=0}^{4} \lambda_{ij} \cdot T^j$$

These Sellmeier models are our best statistical representation of the measured data over the complete measured ranges of wavelength and temperature. All of the following tabulated values for the refractive index have been calculated using this Sellmeier model based on our measured data using the appropriate coefficients in the following tables. Typically the residuals of the fits from measured values are less than the uncertainty respective measurements. These coefficients should not be applied beyond the wavelength and temperatures ranges over which measurements were made.

## 2.1 Calcium Fluoride (CaF$_2$)

Absolute refractive indices of CaF$_2$ were measured over the 0.40 to 5.6 microns wavelength range and over the temperature range from 25 to 300 K for a single test specimen. Refractive index is plotted in Figure 1 and tabulated in Table 1 for selected temperatures and wavelengths. Spectral dispersion is plotted in Figure 2 and tabulated in Table 2. Thermo-optic coefficient is plotted in Figure 3 and tabulated in Table 3. Uncertainties in measured refractive index appear in Table 4. Coefficients for the three term Sellmeier model with 4th order temperature dependence are given in Table 5.

## 2.2 Infrasil 301 measured data

Absolute refractive indices of Infrasil 301 were measured over the 0.40 to 3.6 microns wavelength range and over the temperature range from 35 to 300 K for a single test specimen. Refractive index is plotted in Figure 4 and tabulated in Table 6 for selected temperatures and wavelengths. Spectral dispersion is plotted in Figure 5 and tabulated in Table 7. Thermo-optic coefficient is plotted in Figure 6 and tabulated in Table 8. Uncertainties in measured refractive index appear in Table 9. Coefficients for the three term Sellmeier model with 4th order temperature dependence are given in Table 10.

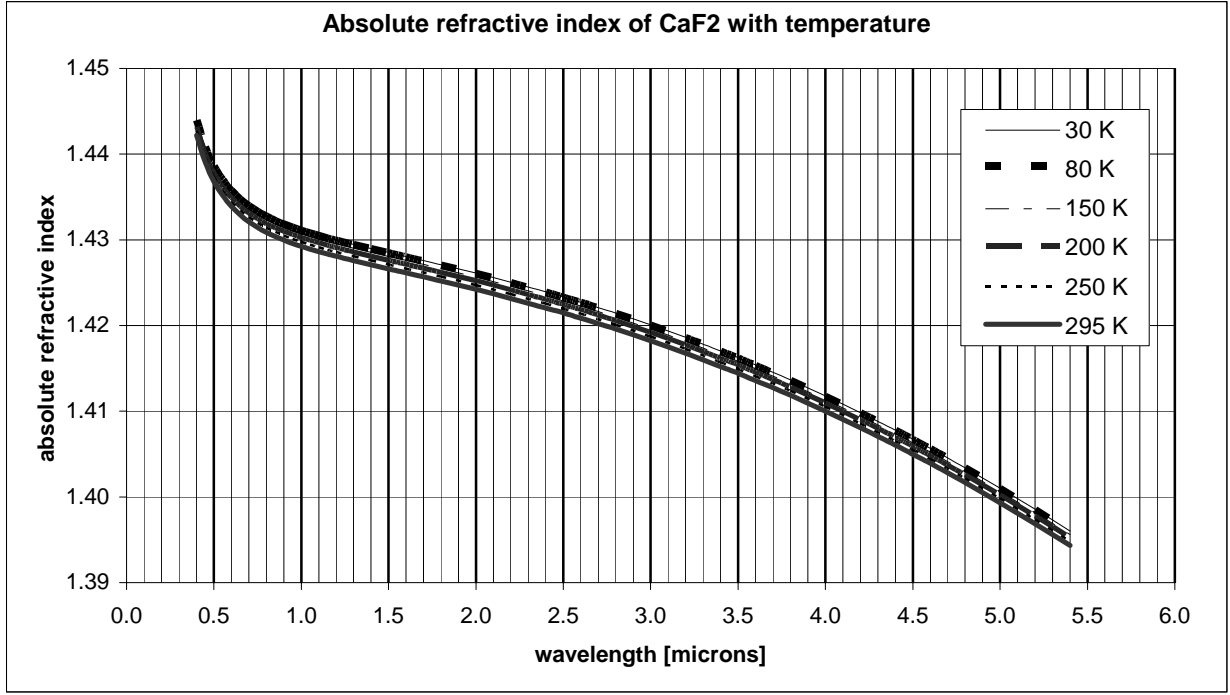

Figure 1 – measured absolute refractive index of CaF$_2$ as a function of wavelength for selected temperatures

Table 1 – measured absolute refractive index of CaF$_2$ for selected wavelengths and temperatures

| Wavelength | 30 K | 40 K | 50 K | 60 K | 70 K | 80 K | 90 K | 100 K | 150 K | 200 K | 250 K | 295 K |
|---|---|---|---|---|---|---|---|---|---|---|---|---|
| 0.40 microns | 1.44401 | 1.44401 | 1.44401 | 1.44399 | 1.44398 | 1.44395 | 1.44392 | 1.44388 | 1.44358 | 1.44316 | 1.44269 | 1.44221 |
| 0.50 microns | 1.43867 | 1.43867 | 1.43866 | 1.43865 | 1.43864 | 1.43861 | 1.43858 | 1.43854 | 1.43825 | 1.43782 | 1.43732 | 1.43683 |
| 0.60 microns | 1.43579 | 1.43578 | 1.43578 | 1.43577 | 1.43575 | 1.43573 | 1.43570 | 1.43566 | 1.43536 | 1.43492 | 1.43441 | 1.43392 |
| 0.70 microns | 1.43401 | 1.43400 | 1.43400 | 1.43399 | 1.43397 | 1.43394 | 1.43391 | 1.43387 | 1.43357 | 1.43313 | 1.43262 | 1.43212 |
| 0.80 microns | 1.43278 | 1.43278 | 1.43278 | 1.43276 | 1.43275 | 1.43272 | 1.43269 | 1.43264 | 1.43234 | 1.43190 | 1.43139 | 1.43089 |
| 0.90 microns | 1.43187 | 1.43187 | 1.43187 | 1.43186 | 1.43184 | 1.43181 | 1.43178 | 1.43173 | 1.43142 | 1.43099 | 1.43047 | 1.42998 |
| 1.00 microns | 1.43115 | 1.43115 | 1.43114 | 1.43113 | 1.43111 | 1.43108 | 1.43105 | 1.43101 | 1.43070 | 1.43026 | 1.42975 | 1.42925 |
| 1.50 microns | 1.42853 | 1.42853 | 1.42852 | 1.42851 | 1.42849 | 1.42846 | 1.42842 | 1.42838 | 1.42807 | 1.42763 | 1.42712 | 1.42662 |
| 2.00 microns | 1.42611 | 1.42611 | 1.42611 | 1.42610 | 1.42607 | 1.42605 | 1.42601 | 1.42597 | 1.42565 | 1.42522 | 1.42472 | 1.42422 |
| 2.50 microns | 1.42334 | 1.42334 | 1.42334 | 1.42332 | 1.42330 | 1.42327 | 1.42324 | 1.42319 | 1.42289 | 1.42246 | 1.42196 | 1.42147 |
| 3.00 microns | 1.42006 | 1.42007 | 1.42006 | 1.42005 | 1.42003 | 1.42000 | 1.41996 | 1.41992 | 1.41962 | 1.41920 | 1.41870 | 1.41821 |
| 3.50 microns | 1.41623 | 1.41623 | 1.41622 | 1.41621 | 1.41619 | 1.41616 | 1.41613 | 1.41608 | 1.41578 | 1.41537 | 1.41488 | 1.41440 |
| 4.00 microns | 1.41179 | 1.41179 | 1.41179 | 1.41178 | 1.41176 | 1.41173 | 1.41169 | 1.41165 | 1.41136 | 1.41096 | 1.41047 | 1.41001 |
| 4.50 microns | 1.40674 | 1.40674 | 1.40673 | 1.40672 | 1.40670 | 1.40668 | 1.40664 | 1.40660 | 1.40632 | 1.40592 | 1.40545 | 1.40499 |
| 5.00 microns | 1.40104 | 1.40104 | 1.40104 | 1.40103 | 1.40101 | 1.40098 | 1.40095 | 1.40091 | 1.40063 | 1.40024 | 1.39978 | 1.39933 |
| 5.50 microns | 1.39467 | 1.39467 | 1.39467 | 1.39466 | 1.39464 | 1.39462 | 1.39459 | 1.39455 | 1.39427 | 1.39390 | 1.39345 | 1.39301 |

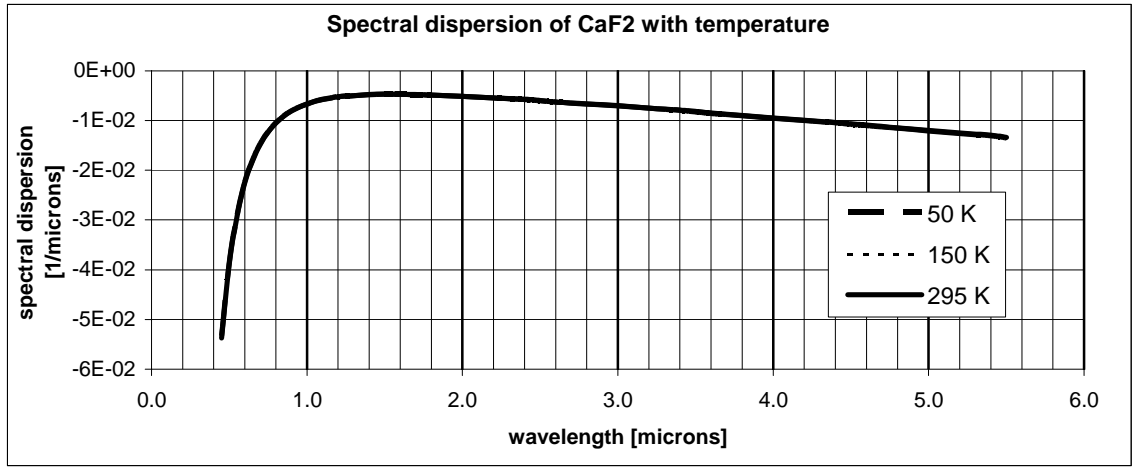

Figure 2 – measured spectral dispersion (dn/dλ) of CaF$_2$ as a function of wavelength for selected temperatures

Table 2 – measured spectral dispersion (dn/dλ) of CaF$_2$ for selected wavelengths and temperatures

| Wavelength | 30 K | 40 K | 50 K | 60 K | 70 K | 80 K | 90 K | 100 K | 150 K | 200 K | 250 K | 295 K |
|---|---|---|---|---|---|---|---|---|---|---|---|---|
| 0.50 microns | -3.9E-02 | -3.9E-02 | -3.9E-02 | -3.9E-02 | -3.9E-02 | -3.9E-02 | -3.9E-02 | -3.9E-02 | -3.9E-02 | -3.9E-02 | -3.9E-02 | -3.9E-02 |
| 0.60 microns | -2.2E-02 | -2.2E-02 | -2.2E-02 | -2.2E-02 | -2.2E-02 | -2.2E-02 | -2.2E-02 | -2.2E-02 | -2.2E-02 | -2.2E-02 | -2.2E-02 | -2.2E-02 |
| 0.70 microns | -1.5E-02 | -1.5E-02 | -1.5E-02 | -1.5E-02 | -1.5E-02 | -1.5E-02 | -1.5E-02 | -1.5E-02 | -1.5E-02 | -1.5E-02 | -1.5E-02 | -1.5E-02 |
| 0.80 microns | -1.0E-02 | -1.0E-02 | -1.0E-02 | -1.0E-02 | -1.0E-02 | -1.0E-02 | -1.0E-02 | -1.0E-02 | -1.0E-02 | -1.1E-02 | -1.1E-02 | -1.1E-02 |
| 0.90 microns | -8.0E-03 | -8.0E-03 | -8.0E-03 | -8.0E-03 | -8.0E-03 | -8.0E-03 | -8.0E-03 | -8.0E-03 | -8.1E-03 | -8.1E-03 | -8.1E-03 | -8.1E-03 |
| 1.00 microns | -6.7E-03 | -6.7E-03 | -6.7E-03 | -6.7E-03 | -6.7E-03 | -6.7E-03 | -6.7E-03 | -6.7E-03 | -6.7E-03 | -6.7E-03 | -6.7E-03 | -6.7E-03 |
| 1.50 microns | -4.7E-03 | -4.7E-03 | -4.7E-03 | -4.7E-03 | -4.7E-03 | -4.7E-03 | -4.7E-03 | -4.7E-03 | -4.7E-03 | -4.7E-03 | -4.7E-03 | -4.7E-03 |
| 2.00 microns | -5.1E-03 | -5.1E-03 | -5.1E-03 | -5.1E-03 | -5.1E-03 | -5.1E-03 | -5.1E-03 | -5.1E-03 | -5.1E-03 | -5.1E-03 | -5.1E-03 | -5.1E-03 |
| 2.50 microns | -6.0E-03 | -6.0E-03 | -6.0E-03 | -6.0E-03 | -6.0E-03 | -6.0E-03 | -6.0E-03 | -6.0E-03 | -6.0E-03 | -6.0E-03 | -6.0E-03 | -6.0E-03 |
| 3.00 microns | -7.1E-03 | -7.1E-03 | -7.1E-03 | -7.1E-03 | -7.1E-03 | -7.1E-03 | -7.0E-03 | -7.0E-03 | -7.0E-03 | -7.0E-03 | -7.0E-03 | -7.0E-03 |
| 3.50 microns | -8.2E-03 | -8.3E-03 | -8.3E-03 | -8.3E-03 | -8.3E-03 | -8.3E-03 | -8.3E-03 | -8.3E-03 | -8.2E-03 | -8.2E-03 | -8.2E-03 | -8.2E-03 |
| 4.00 microns | -9.5E-03 | -9.5E-03 | -9.5E-03 | -9.5E-03 | -9.5E-03 | -9.5E-03 | -9.5E-03 | -9.5E-03 | -9.5E-03 | -9.5E-03 | -9.5E-03 | -9.5E-03 |
| 4.50 microns | -1.1E-02 | -1.1E-02 | -1.1E-02 | -1.1E-02 | -1.1E-02 | -1.1E-02 | -1.1E-02 | -1.1E-02 | -1.1E-02 | -1.1E-02 | -1.1E-02 | -1.1E-02 |
| 5.00 microns | -1.2E-02 | -1.2E-02 | -1.2E-02 | -1.2E-02 | -1.2E-02 | -1.2E-02 | -1.2E-02 | -1.2E-02 | -1.2E-02 | -1.2E-02 | -1.2E-02 | -1.2E-02 |
| 5.50 microns | -1.4E-02 | -1.4E-02 | -1.4E-02 | -1.3E-02 | -1.3E-02 | -1.3E-02 | -1.3E-02 | -1.3E-02 | -1.3E-02 | -1.3E-02 | -1.3E-02 | -1.3E-02 |

Table 3 – measured thermo-optic coefficient (dn/dT) of CaF$_2$ for selected wavelengths and temperatures

| wavelength | 30 K | 40 K | 50 K | 60 K | 70 K | 80 K | 90 K | 100 K | 150 K | 200 K | 250 K | 295 K |
|---|---|---|---|---|---|---|---|---|---|---|---|---|
| 0.40 microns | 3.5E-07 | -2.9E-07 | -9.3E-07 | -1.6E-06 | -2.2E-06 | -2.9E-06 | -3.5E-06 | -4.2E-06 | -7.8E-06 | -8.9E-06 | -1.0E-05 | -1.1E-05 |
| 0.50 microns | 3.8E-07 | -2.8E-07 | -9.4E-07 | -1.6E-06 | -2.3E-06 | -2.9E-06 | -3.6E-06 | -4.3E-06 | -8.0E-06 | -9.2E-06 | -1.0E-05 | -1.1E-05 |
| 0.60 microns | 5.7E-07 | -1.4E-07 | -8.4E-07 | -1.5E-06 | -2.2E-06 | -2.9E-06 | -3.6E-06 | -4.4E-06 | -8.1E-06 | -9.3E-06 | -1.0E-05 | -1.2E-05 |
| 0.70 microns | 5.3E-07 | -1.7E-07 | -8.7E-07 | -1.6E-06 | -2.3E-06 | -3.0E-06 | -3.7E-06 | -4.4E-06 | -8.1E-06 | -9.3E-06 | -1.1E-05 | -1.2E-05 |
| 0.80 microns | 6.0E-07 | -1.2E-07 | -8.4E-07 | -1.6E-06 | -2.3E-06 | -3.0E-06 | -3.7E-06 | -4.5E-06 | -8.3E-06 | -9.4E-06 | -1.1E-05 | -1.2E-05 |
| 0.90 microns | 6.1E-07 | -1.2E-07 | -8.5E-07 | -1.6E-06 | -2.3E-06 | -3.0E-06 | -3.7E-06 | -4.6E-06 | -8.3E-06 | -9.4E-06 | -1.1E-05 | -1.2E-05 |
| 1.00 microns | 5.8E-07 | -1.4E-07 | -8.7E-07 | -1.6E-06 | -2.3E-06 | -3.0E-06 | -3.8E-06 | -4.6E-06 | -8.2E-06 | -9.4E-06 | -1.1E-05 | -1.2E-05 |
| 1.50 microns | 3.4E-07 | -3.6E-07 | -1.1E-06 | -1.7E-06 | -2.4E-06 | -3.1E-06 | -3.8E-06 | -4.6E-06 | -8.1E-06 | -9.4E-06 | -1.1E-05 | -1.2E-05 |
| 2.00 microns | 2.9E-07 | -3.9E-07 | -1.1E-06 | -1.7E-06 | -2.4E-06 | -3.1E-06 | -3.8E-06 | -4.6E-06 | -8.0E-06 | -9.3E-06 | -1.1E-05 | -1.2E-05 |
| 2.50 microns | 3.5E-07 | -3.5E-07 | -1.0E-06 | -1.7E-06 | -2.4E-06 | -3.1E-06 | -3.8E-06 | -4.6E-06 | -7.7E-06 | -9.1E-06 | -1.0E-05 | -1.2E-05 |
| 3.00 microns | 2.1E-07 | -4.4E-07 | -1.1E-06 | -1.8E-06 | -2.4E-06 | -3.1E-06 | -3.7E-06 | -4.5E-06 | -7.8E-06 | -9.1E-06 | -1.0E-05 | -1.1E-05 |
| 3.50 microns | 2.4E-07 | -4.1E-07 | -1.1E-06 | -1.7E-06 | -2.4E-06 | -3.0E-06 | -3.7E-06 | -4.4E-06 | -7.5E-06 | -8.9E-06 | -1.0E-05 | -1.1E-05 |
| 4.00 microns | 3.2E-07 | -3.4E-07 | -1.0E-06 | -1.7E-06 | -2.3E-06 | -3.0E-06 | -3.6E-06 | -4.4E-06 | -7.4E-06 | -8.7E-06 | -1.0E-05 | -1.1E-05 |
| 4.50 microns | 5.1E-07 | -1.9E-07 | -8.9E-07 | -1.6E-06 | -2.3E-06 | -3.0E-06 | -3.7E-06 | -4.5E-06 | -7.4E-06 | -8.6E-06 | -9.8E-06 | -1.1E-05 |
| 5.00 microns | 7.8E-07 | 4.7E-08 | -6.9E-07 | -1.4E-06 | -2.2E-06 | -2.9E-06 | -3.6E-06 | -4.4E-06 | -7.1E-06 | -8.4E-06 | -9.6E-06 | -1.1E-05 |
| 5.50 microns | 8.3E-07 | 1.1E-07 | -6.1E-07 | -1.3E-06 | -2.1E-06 | -2.8E-06 | -3.5E-06 | -4.3E-06 | -7.0E-06 | -8.1E-06 | -9.3E-06 | -1.0E-05 |

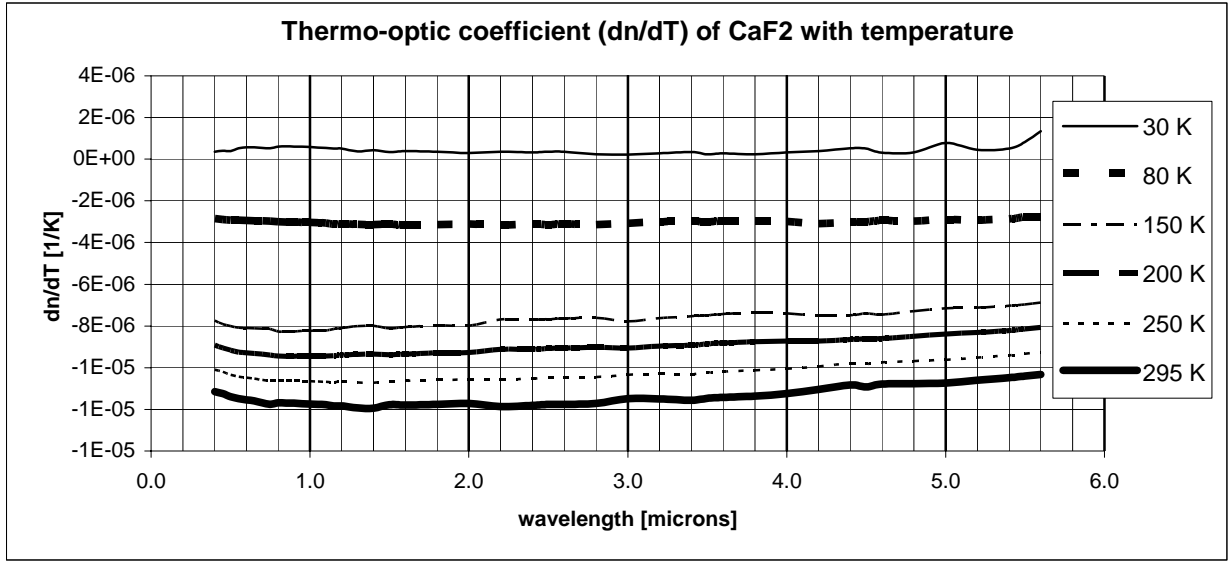

Figure 3 – measured thermo-optic coefficient (dn/dT) of $CaF_2$ as a function of wavelength for selected temperatures

Table 4 – uncertainty in absolute refractive index measurements of $CaF_2$ for selected wavelengths and temperatures

| wavelength | 30 K | 75 K | 100 K | 200 K | 295 K |
|---|---|---|---|---|---|
| 0.50 microns | 1.7E-05 | 1.7E-05 | 1.8E-05 | 1.8E-05 | 1.7E-05 |
| 1.0 microns | 9.2E-06 | 9.6E-06 | 1.0E-05 | 1.0E-05 | 9.7E-06 |
| 2.0 microns | 8.6E-06 | 9.0E-06 | 9.6E-06 | 9.7E-06 | 9.0E-06 |
| 4.0 microns | 1.1E-05 | 1.1E-05 | 1.2E-05 | 1.2E-05 | 1.1E-05 |
| 5.5 microns | 1.3E-05 | 1.3E-05 | 1.4E-05 | 1.4E-05 | 1.3E-05 |

Table 5 – coefficients for the temperature-dependent Sellmeier fit of the refractive index of $CaF_2$ measured in CHARMS; average absolute residual of the fit from the measured data is $6 \times 10^{-6}$

| Coefficients for the temperature dependent Sellmeier equation for $CaF_2$ | | | | | |
|---|---|---|---|---|---|
| 25 K ≤ T ≤ 300 K; 0.40 microns ≤ λ ≤ 5.6 microns | | | | | |
| $S_1$ | $S_2$ | $S_3$ | $\lambda_1$ | $\lambda_2$ | $\lambda_3$ |
| **Constant term** 1.04834 | -3.32723E-03 | 3.72693 | 7.94375E-02 | 0.258039 | 34.0169 |
| **T term** -2.21666E-04 | 2.34683E-04 | 1.49844E-02 | -2.20758E-04 | -2.12833E-03 | 6.26867E-02 |
| **$T^2$ term** -6.73446E-06 | 6.55744E-06 | -1.47511E-04 | 2.07862E-06 | 1.20393E-05 | -6.14541E-04 |
| **$T^3$ term** 1.50138E-08 | -1.47028E-08 | 5.54293E-07 | -9.60254E-09 | -3.06973E-08 | 2.31517E-06 |
| **$T^4$ term** -2.77255E-11 | 2.75023E-11 | -7.17298E-10 | 1.31401E-11 | 2.79793E-11 | -2.99638E-09 |

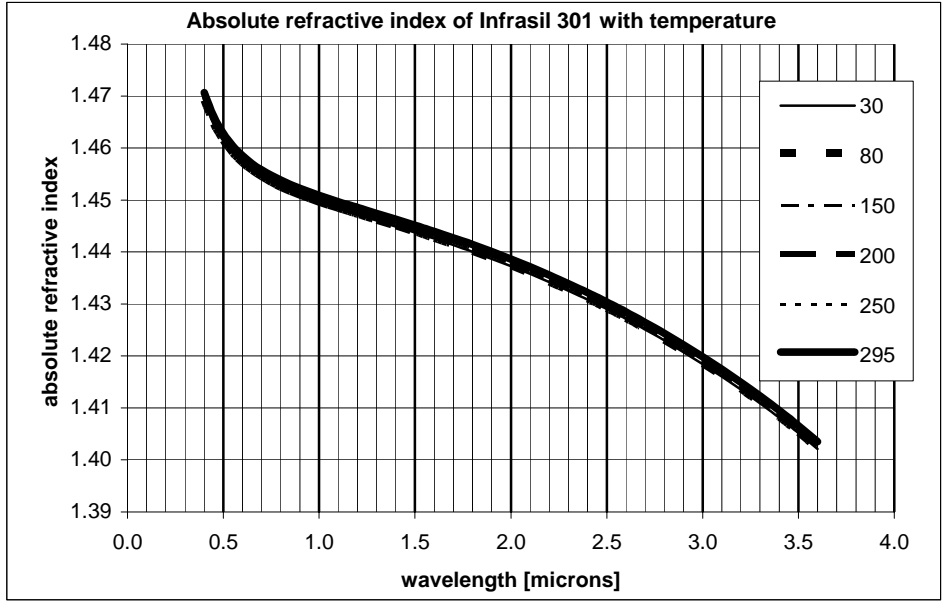

Figure 4 – measured absolute refractive index of Infrasil 301 as a function of wavelength for selected temperatures

Table 6 – measured absolute refractive index of Infrasil 301 for selected wavelengths and temperatures

| Wavelength | 30 K | 40 K | 50 K | 60 K | 70 K | 80 K | 90 K | 100 K | 150 K | 200 K | 250 K | 295 K |
|---|---|---|---|---|---|---|---|---|---|---|---|---|
| 0.40 microns* | 1.46908 | 1.46909 | 1.46911 | 1.46913 | 1.46916 | 1.46919 | 1.46922 | 1.46926 | 1.46951 | 1.46984 | 1.47024 | 1.47064 |
| 0.50 microns | 1.46136 | 1.46138 | 1.46139 | 1.46142 | 1.46144 | 1.46147 | 1.46151 | 1.46154 | 1.46178 | 1.46209 | 1.46246 | 1.46285 |
| 0.60 microns | 1.45710 | 1.45712 | 1.45714 | 1.45716 | 1.45719 | 1.45722 | 1.45725 | 1.45729 | 1.45751 | 1.45781 | 1.45818 | 1.45855 |
| 0.70 microns | 1.45437 | 1.45439 | 1.45441 | 1.45443 | 1.45445 | 1.45448 | 1.45452 | 1.45455 | 1.45478 | 1.45507 | 1.45543 | 1.45579 |
| 0.80 microns | 1.45240 | 1.45242 | 1.45244 | 1.45246 | 1.45249 | 1.45252 | 1.45255 | 1.45258 | 1.45281 | 1.45310 | 1.45345 | 1.45381 |
| 0.90 microns | 1.45085 | 1.45086 | 1.45088 | 1.45090 | 1.45093 | 1.45096 | 1.45099 | 1.45103 | 1.45125 | 1.45153 | 1.45188 | 1.45224 |
| 1.00 microns | 1.44952 | 1.44953 | 1.44955 | 1.44957 | 1.44960 | 1.44963 | 1.44966 | 1.44969 | 1.44991 | 1.45020 | 1.45055 | 1.45090 |
| 1.50 microns | 1.44375 | 1.44376 | 1.44378 | 1.44380 | 1.44383 | 1.44385 | 1.44388 | 1.44392 | 1.44413 | 1.44442 | 1.44476 | 1.44512 |
| 2.00 microns | 1.43726 | 1.43727 | 1.43729 | 1.43731 | 1.43733 | 1.43736 | 1.43739 | 1.43742 | 1.43763 | 1.43792 | 1.43827 | 1.43863 |
| 2.50 microns | 1.42903 | 1.42903 | 1.42905 | 1.42907 | 1.42909 | 1.42911 | 1.42914 | 1.42917 | 1.42938 | 1.42967 | 1.43003 | 1.43039 |
| 3.00 microns | 1.41850 | 1.41851 | 1.41853 | 1.41855 | 1.41857 | 1.41859 | 1.41862 | 1.41866 | 1.41887 | 1.41916 | 1.41952 | 1.41988 |
| 3.50 microns | 1.40514 | 1.40516 | 1.40518 | 1.40520 | 1.40523 | 1.40526 | 1.40529 | 1.40533 | 1.40555 | 1.40584 | 1.40619 | 1.40655 |

* for 0.4 microns, measured index data interpolated to selected temperature is reported; other values are from Sellmeier fit

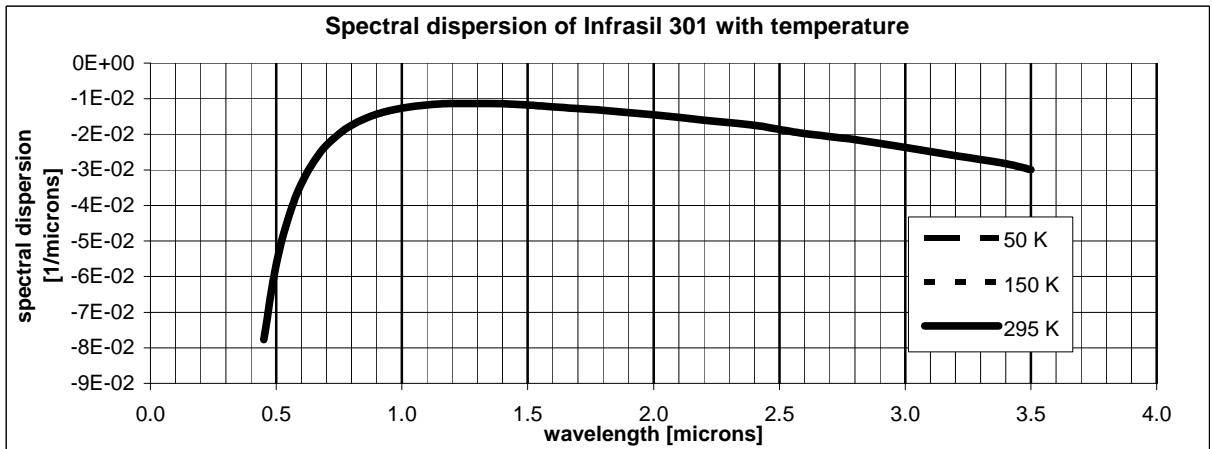

Figure 5 – measured spectral dispersion (dn/dλ) of Infrasil 301 as a function of wavelength for selected temperatures

Table 7 – measured spectral dispersion (dn/dλ) of Infrasil 301 for selected wavelengths and temperatures

| wavelength | 30 K | 40 K | 50 K | 60 K | 70 K | 80 K | 90 K | 100 K | 150 K | 200 K | 250 K | 295 K |
|---|---|---|---|---|---|---|---|---|---|---|---|---|
| 0.50 microns | -5.6E-02 | -5.6E-02 | -5.6E-02 | -5.6E-02 | -5.6E-02 | -5.6E-02 | -5.6E-02 | -5.6E-02 | -5.6E-02 | -5.6E-02 | -5.6E-02 | -5.7E-02 |
| 0.60 microns | -3.4E-02 | -3.4E-02 | -3.4E-02 | -3.4E-02 | -3.4E-02 | -3.4E-02 | -3.4E-02 | -3.4E-02 | -3.4E-02 | -3.4E-02 | -3.4E-02 | -3.4E-02 |
| 0.70 microns | -2.3E-02 | -2.3E-02 | -2.3E-02 | -2.3E-02 | -2.3E-02 | -2.3E-02 | -2.3E-02 | -2.3E-02 | -2.3E-02 | -2.3E-02 | -2.3E-02 | -2.3E-02 |
| 0.80 microns | -1.7E-02 | -1.7E-02 | -1.7E-02 | -1.7E-02 | -1.7E-02 | -1.7E-02 | -1.7E-02 | -1.7E-02 | -1.7E-02 | -1.7E-02 | -1.7E-02 | -1.7E-02 |
| 0.90 microns | -1.4E-02 | -1.4E-02 | -1.4E-02 | -1.4E-02 | -1.4E-02 | -1.4E-02 | -1.4E-02 | -1.4E-02 | -1.4E-02 | -1.4E-02 | -1.4E-02 | -1.4E-02 |
| 1.00 microns | -1.3E-02 | -1.3E-02 | -1.3E-02 | -1.3E-02 | -1.3E-02 | -1.3E-02 | -1.3E-02 | -1.3E-02 | -1.3E-02 | -1.3E-02 | -1.3E-02 | -1.3E-02 |
| 1.50 microns | -1.2E-02 | -1.2E-02 | -1.2E-02 | -1.2E-02 | -1.2E-02 | -1.2E-02 | -1.2E-02 | -1.2E-02 | -1.2E-02 | -1.2E-02 | -1.2E-02 | -1.2E-02 |
| 2.00 microns | -1.5E-02 | -1.5E-02 | -1.5E-02 | -1.5E-02 | -1.5E-02 | -1.5E-02 | -1.5E-02 | -1.5E-02 | -1.5E-02 | -1.5E-02 | -1.5E-02 | -1.5E-02 |
| 2.50 microns | -1.9E-02 | -1.9E-02 | -1.9E-02 | -1.9E-02 | -1.9E-02 | -1.9E-02 | -1.9E-02 | -1.9E-02 | -1.9E-02 | -1.9E-02 | -1.9E-02 | -1.9E-02 |
| 3.00 microns | -2.4E-02 | -2.4E-02 | -2.4E-02 | -2.4E-02 | -2.4E-02 | -2.4E-02 | -2.4E-02 | -2.4E-02 | -2.4E-02 | -2.4E-02 | -2.4E-02 | -2.4E-02 |
| 3.50 microns | -3.0E-02 | -3.0E-02 | -3.0E-02 | -3.0E-02 | -3.0E-02 | -3.0E-02 | -3.0E-02 | -3.0E-02 | -3.0E-02 | -3.0E-02 | -3.0E-02 | -3.0E-02 |

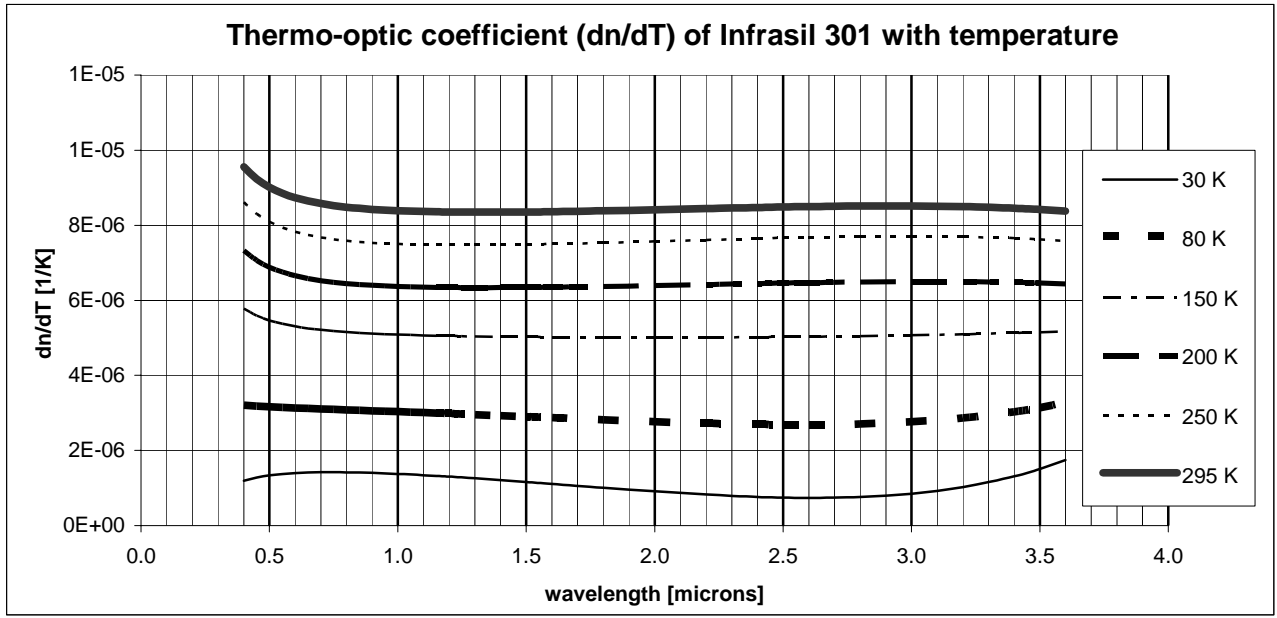

Figure 6 – measured thermo-optic coefficient (dn/dT) of Infrasil 301 as a function of wavelength for selected temperatures

Table 8 – measured thermo-optic coefficient (dn/dT) of Infrasil 301 for selected wavelengths and temperatures

| wavelength | 30 K | 40 K | 50 K | 60 K | 70 K | 80 K | 90 K | 100 K | 150 K | 200 K | 250 K | 295 K |
|---|---|---|---|---|---|---|---|---|---|---|---|---|
| 0.40 microns | 1.2E-06 | 1.6E-06 | 2.0E-06 | 2.4E-06 | 2.8E-06 | 3.2E-06 | 3.6E-06 | 4.0E-06 | 5.8E-06 | 7.3E-06 | 8.6E-06 | 9.6E-06 |
| 0.50 microns | 1.3E-06 | 1.7E-06 | 2.1E-06 | 2.4E-06 | 2.8E-06 | 3.2E-06 | 3.5E-06 | 3.9E-06 | 5.5E-06 | 6.9E-06 | 8.1E-06 | 9.0E-06 |
| 0.60 microns | 1.4E-06 | 1.8E-06 | 2.1E-06 | 2.5E-06 | 2.8E-06 | 3.1E-06 | 3.5E-06 | 3.8E-06 | 5.3E-06 | 6.7E-06 | 7.8E-06 | 8.7E-06 |
| 0.70 microns | 1.4E-06 | 1.8E-06 | 2.1E-06 | 2.4E-06 | 2.8E-06 | 3.1E-06 | 3.4E-06 | 3.8E-06 | 5.2E-06 | 6.5E-06 | 7.7E-06 | 8.6E-06 |
| 0.80 microns | 1.4E-06 | 1.8E-06 | 2.1E-06 | 2.4E-06 | 2.8E-06 | 3.1E-06 | 3.4E-06 | 3.7E-06 | 5.2E-06 | 6.4E-06 | 7.6E-06 | 8.5E-06 |
| 0.90 microns | 1.4E-06 | 1.7E-06 | 2.1E-06 | 2.4E-06 | 2.7E-06 | 3.1E-06 | 3.4E-06 | 3.7E-06 | 5.1E-06 | 6.4E-06 | 7.5E-06 | 8.4E-06 |
| 1.00 microns | 1.4E-06 | 1.7E-06 | 2.1E-06 | 2.4E-06 | 2.7E-06 | 3.0E-06 | 3.3E-06 | 3.7E-06 | 5.1E-06 | 6.4E-06 | 7.5E-06 | 8.4E-06 |
| 1.50 microns | 1.2E-06 | 1.5E-06 | 1.9E-06 | 2.2E-06 | 2.6E-06 | 2.9E-06 | 3.2E-06 | 3.6E-06 | 5.0E-06 | 6.3E-06 | 7.5E-06 | 8.4E-06 |
| 2.00 microns | 9.1E-07 | 1.3E-06 | 1.7E-06 | 2.1E-06 | 2.4E-06 | 2.8E-06 | 3.1E-06 | 3.5E-06 | 5.0E-06 | 6.4E-06 | 7.6E-06 | 8.4E-06 |
| 2.50 microns | 7.5E-07 | 1.2E-06 | 1.6E-06 | 1.9E-06 | 2.3E-06 | 2.7E-06 | 3.0E-06 | 3.4E-06 | 5.0E-06 | 6.5E-06 | 7.7E-06 | 8.5E-06 |
| 3.00 microns | 8.5E-07 | 1.3E-06 | 1.7E-06 | 2.0E-06 | 2.4E-06 | 2.8E-06 | 3.1E-06 | 3.5E-06 | 5.1E-06 | 6.5E-06 | 7.7E-06 | 8.5E-06 |
| 3.50 microns | 1.5E-06 | 1.9E-06 | 2.2E-06 | 2.5E-06 | 2.8E-06 | 3.1E-06 | 3.4E-06 | 3.8E-06 | 5.1E-06 | 6.5E-06 | 7.6E-06 | 8.4E-06 |

Table 9 – uncertainty in absolute refractive index measurements of Infrasil 301 for selected wavelengths and temperatures

| wavelength | 30 K | 75 K | 100 K | 200 K | 295 K |
|---|---|---|---|---|---|
| 0.50 microns | 2.4E-05 | 2.4E-05 | 2.4E-05 | 2.4E-05 | 2.4E-05 |
| 1.0 microns | 1.3E-05 | 1.3E-05 | 1.3E-05 | 1.3E-05 | 1.3E-05 |
| 2.0 microns | 1.4E-05 | 1.4E-05 | 1.4E-05 | 1.4E-05 | 1.4E-05 |
| 3.0 microns | 2.0E-05 | 2.0E-05 | 2.0E-05 | 2.0E-05 | 2.0E-05 |

Table 10 – coefficients for the temperature-dependent Sellmeier fit of the refractive index of Infrasil 301 measured in CHARMS; average absolute residual of the fit from the measured data is $1 \times 10^{-5}$

| **Coefficients for the temperature dependent Sellmeier equation for Infrasil 301** $35\ K \leq T \leq 300\ K;\ 0.50\ \text{microns} \leq \lambda \leq 3.6\ \text{microns}$ | | | | | | |
|---|---|---|---|---|---|---|
|  | $S_1$ | $S_2$ | $S_3$ | $\lambda_1$ | $\lambda_2$ | $\lambda_3$ |
| **Constant term** | 0.105962 | 0.995429 | 0.865120 | 4.500743E-03 | 9.383735E-02 | 9.757183 |
| **T term** | 9.359142E-06 | -7.973196E-06 | 3.731950E-04 | -2.825065E-04 | -1.374171E-06 | 1.864621E-03 |
| **$T^2$ term** | 4.941067E-08 | 1.006343E-09 | -2.010347E-06 | 3.136868E-06 | 1.316037E-08 | -1.058414E-05 |
| **$T^3$ term** | 4.890163E-11 | -8.694712E-11 | 2.708606E-09 | -1.121499E-08 | 1.252909E-11 | 1.730321E-08 |
| **$T^4$ term** | 1.492126E-13 | -1.220612E-13 | 1.679976E-12 | 1.236514E-11 | -4.641280E-14 | 1.719396E-12 |

## 3. COMPARISON WITH REFRACTIVE INDEX VALUES FROM THE LITERATURE

**3.1 CaF$_2$**
There is very little data in the literature, especially for cryogenic temperatures, to which to compare our measurements using CHARMS. There have been several previous investigations of the refractive index of CaF$_2$, primarily at room temperature. Tropf[10] generated temperature-dependent Sellmeier models based on room temperature refractive index measurements of Malitson[11] and measurements of dn/dT by Feldman, et al[12] over the temperature range 93 K to 473 K. While we have previously found that our measurements agree well with Tropf's models for other materials to within respective measurement uncertainties[5], initially, our measurements appeared to be vastly discrepant with Tropf's model for CaF$_2$, even at room temperature, monotonically departing from his model from a small value to just more than one part in the third decimal place of index near 3.39 microns, the long wavelength end of that model's range of validity. Meanwhile, our room temperature measurements agree reasonably well with both Malitson and Feldman (upon which Tropf's model is based) as well as with more recent data from Daimon and Masumura[13] (see Figure 7). Note that these are comparisons of absolute refractive index, that is, not with respect to air, as cryogenic measurements have invariably involved providing a vacuum around the sample.

After discovering the discrepancy of our results with the well-trusted Tropf model, we deduced that there must be a typo in Tropf's published coefficients. We contacted Tropf about this possibility, and he generously revisited his model's coefficients and determined that there was indeed a typo in his published results. He developed a revised set of coefficients for the previously used data and provided them to us for comparison with our measured index data.[14] Upon applying those coefficients, we found reasonable agreement between our measured indices and those predicted by Tropf's revised model (Figure 8). Our values are typically about $5 \times 10^{-5}$ lower than the revised Tropf model. As crystal growth techniques have changed considerably in the last 30 or 40 years, differences in index could easily be attributable to interspecimen variability between the 2006 vintage specimen we measured and 1963 and 1979 vintage specimens measured by investigators whose data led to Tropf's model. We are eager to study interspecimen variability for this popular infrared optical material for in the modern era.

One other set of recent measurements at cryogenic temperatures has been made by Yamamuro et al[15] from room temperature down to 80 K from 0.35 to 3.3 microns to an uncertainty of +/-1.8 x 10$^{-4}$. Our agreement with Yamamuro is everywhere within $1 \times 10^{-4}$ except at 3.3 microns where it is about $2 \times 10^{-4}$.

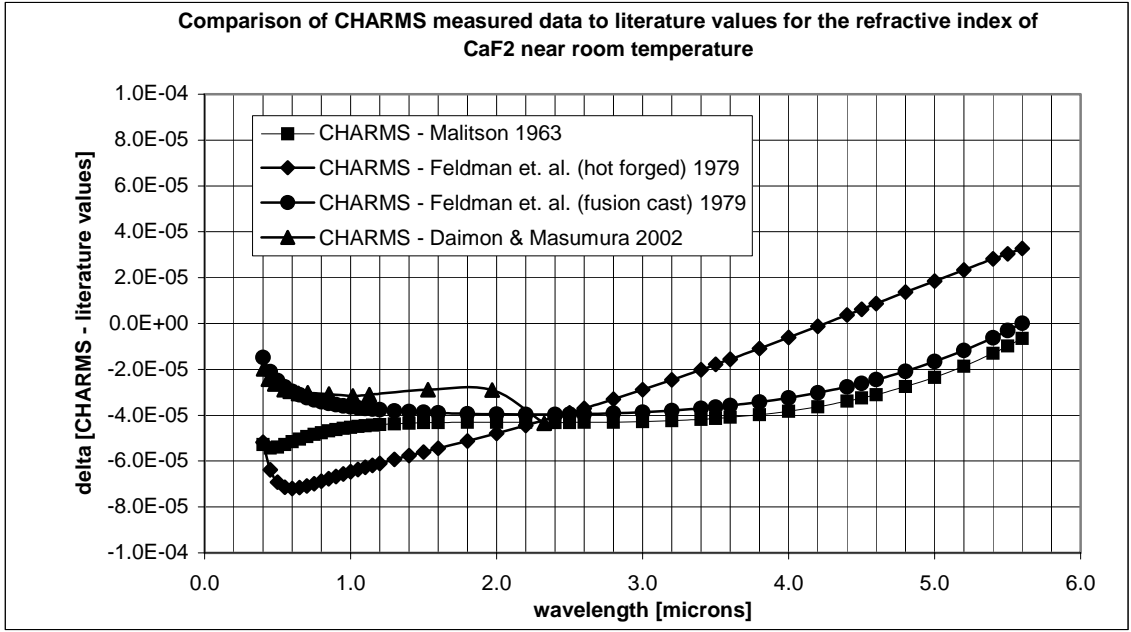

Figure 7 – agreement between CHARMS measured values and various literature values for refractive index of $CaF_2$ at room temperature (excluding Tropf model)

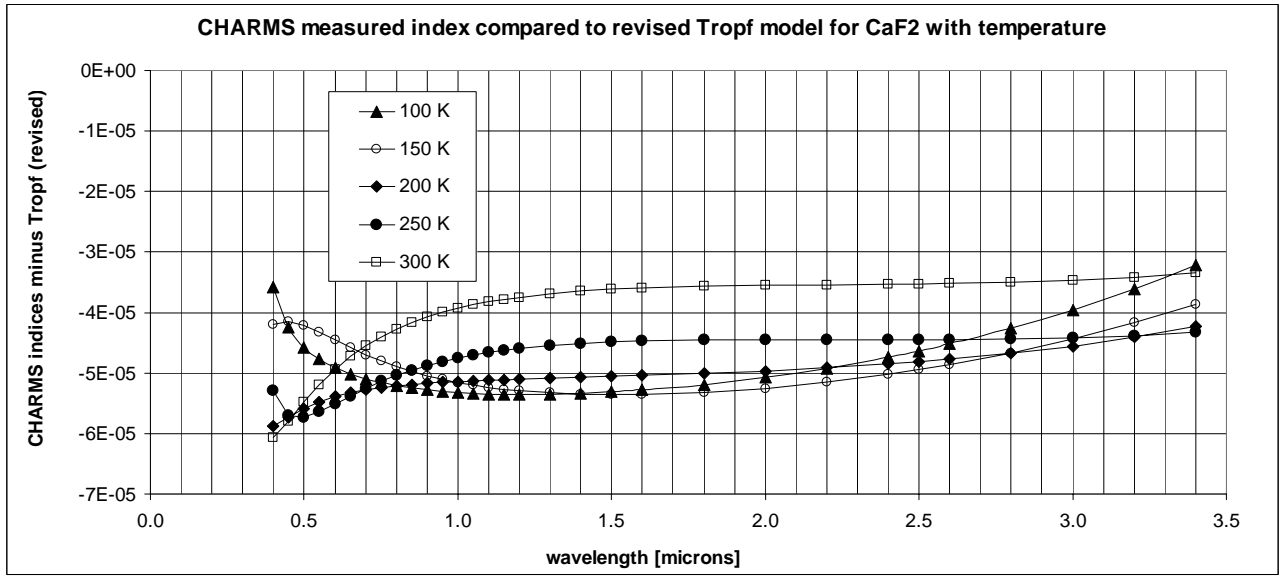

Figure 8 – agreement between CHARMS fitted measured index values and values from revised Tropf model for refractive index of $CaF_2$ at for room temperature

### 3.2 Infrasil 301

Cryogenic index data in the literature for fused silica is surprisingly scarce given the history of the material and its ubiquitous use. However, this data can be found for an unstated type of fused silica from Yamamuro et al[15] and from our own cryogenic measurements of Corning 7980 fused silica two years ago.[6] In this section, we compare our measurements of Infrasil 301 to those two sets of results for low temperatures and to room temperature data from the manufacturer, Heraeus Quarzglas.[16]

Heraeus provides a dispersion equation valid at 293 K and ambient pressure in air with a stated accuracy of +/-3 x $10^{-5}$ but does not state the wavelength range over which the dispersion equation is valid, nor typical values for interspecimen variability. For comparison we convert the manufacturer's index data to vacuum by multiplying by the index of refraction of air (roughly 1.00027). Our measurements agree well with this dispersion equation in the visible and NIR, but begin to depart as we approach the absorption feature at 3.6 microns (Figure 9). This departure is likely due primarily due to increased measurement uncertainty of the material at wavelength in the vicinity of the material's dominant absorption feature at 2.73 microns.

As for cryogenic behavior of the material, absolute measurements by Yamamuro of an unspecified type of fused silica are consistently higher than our CHARMS measurements of Infrasil by from $1.8 \times 10^{-4}$ at 0.436 microns and 82 K to $2.7 \times 10^{-4}$ at 1.530 microns near room temperature – mostly within Yamamuro's measurement uncertainty of $+/-1.8 \times 10^{-4}$. Next, we compare our previous measurements of Corning 7980 fused silica with our current measurements of Infrasil 301. At room temperature, manufacturers' dispersion equations for index for each material show that across the visible wavelength range, Infrasil 301 has an index which is higher than that of Corning 7980 by $1.0 \times 10^{-4}$. The CHARMS measurements of the two materials bear this out nearly exactly (Figure 10). The difference in spectral character of the two materials in the infrared likely depends mostly on different purity levels of the hydroxl molecule OH responsible for the absorption dip at 2.73 microns, which is suppressed in Infrasil by its production process.

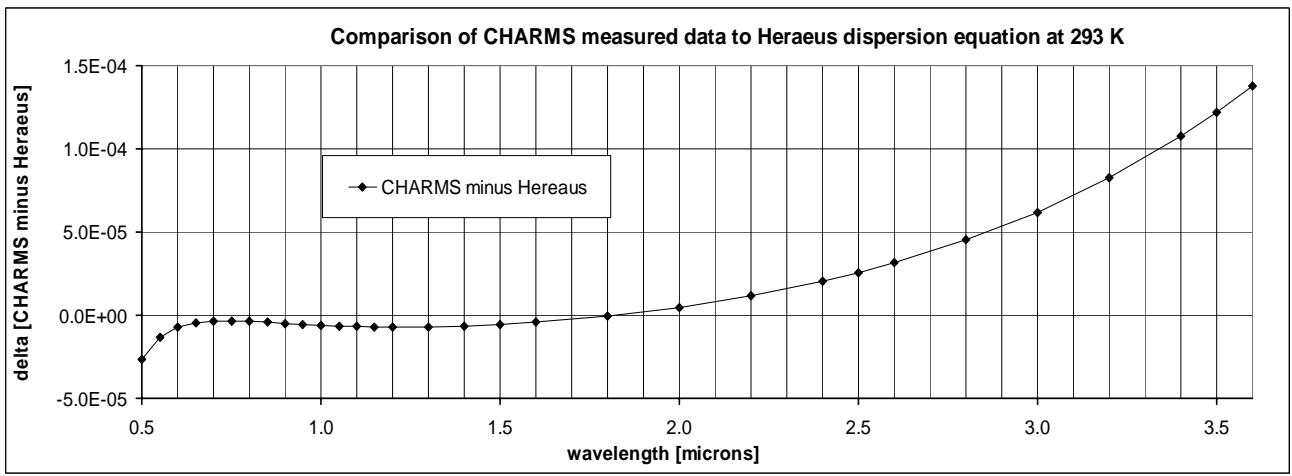

Figure 9 – agreement between CHARMS measured refractive index values and values calculated using Heraeus's dispersion equation for Infrasil 301 at room temperature corrected to vacuum

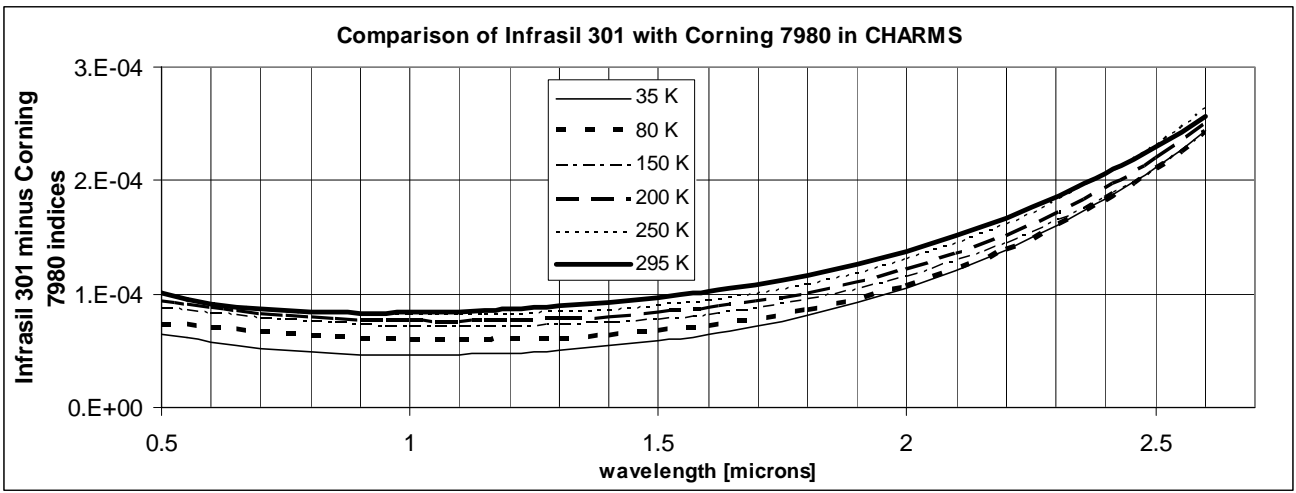

Figure 10 – comparison of measured refractive indices at low temperatures for Infrasil 301 and Corning 7980 using CHARMS

## 4. CONCLUSION

Using the CHARMS facility at NASA GSFC, we have measured the cryogenic refractive index of single crystal $CaF_2$ from 0.40 to 5.6 μm and from 25 to 300K and Heraeus Infrasil 301 from 0.40 to 3.6 μm and from 35 to 300K with high accuracy. We have also examined their spectral dispersion and thermo-optic coefficients, and have derived temperature-dependent Sellmeier models from which refractive index may be calculated for any wavelength and temperature within the stated ranges of each model.